\documentclass[aps,twocolumn,superscriptaddress,notitlepage,longbibliography]{revtex4-1}
\usepackage{tikz}  
\usetikzlibrary{shapes,arrows,positioning,automata,backgrounds,calc,er,patterns}

\DeclareMathAlphabet\mathbfcal{OMS}{cmsy}{b}{n}
\usepackage{graphicx}
\usepackage{dcolumn}
\usepackage{bm}
\usepackage{amsmath}
\usepackage{amssymb}
\usepackage{amsthm}
\usepackage{amsfonts}
\usepackage{enumerate}
\usepackage{bbm}
\usepackage{array}
\newcolumntype{P}[1]{>{\centering\arraybackslash}p{#1}}
\usepackage{float}

\renewcommand{\vec}{\mathbf}
\newcommand{\bra}[1]{\langle #1 \vert}
\newcommand{\ket}[1]{\vert #1 \rangle}

\definecolor{purple}{rgb}{.6, 0., 0.6}

\usepackage{braket}

\usepackage[colorlinks=true,urlcolor=blue,citecolor=blue]{hyperref}
\usepackage{cleveref}

\begin{document}

\title{Entanglement-limited linear response in fermionic systems}

\author{Hadi Cheraghi}
\address{Computational Physics Laboratory, Physics Unit, Faculty of Engineering and
Natural Sciences, Tampere University, P.O. Box 692, FI-33014 Tampere, Finland}
\address{Helsinki Institute of Physics P.O. Box 64, FI-00014, Finland}

\author{Ali G. Moghaddam}
\address{Department of Applied Physics, Aalto University, 02150 Espoo, Finland}
\address{Computational Physics Laboratory, Physics Unit, Faculty of Engineering and
Natural Sciences, Tampere University, P.O. Box 692, FI-33014 Tampere, Finland}
\address{Helsinki Institute of Physics P.O. Box 64, FI-00014, Finland}

\author{Teemu Ojanen}
\address{Computational Physics Laboratory, Physics Unit, Faculty of Engineering and
Natural Sciences, Tampere University, P.O. Box 692, FI-33014 Tampere, Finland}
\address{Helsinki Institute of Physics P.O. Box 64, FI-00014, Finland}

\begin{abstract}
We propose a general connection between entanglement-entropy scaling laws and the linear response functions of particle-conserving fermionic systems in their ground state. Specifically, we show that the response to perturbations coupled to the particle number within a finite region exhibits the same size scaling as the entanglement entropy of that region. We explicitly verify this scaling in free-fermion systems that display area-law, volume-law, and critical forms of entanglement. The resulting entanglement-governed scaling of response functions leads to unexpected physical consequences. For instance, contrary to conventional expectations, the energy absorption rate and particle-number fluctuations in gapped systems scale with the boundary of the perturbed region rather than with its volume. Our work thus establishes a direct link between linear-response properties and many-body entanglement.
\end{abstract}
\maketitle

\section{Introduction}

Despite the central role of entanglement in quantum mechanics, it played a surprisingly modest part in the development of quantum many-body and field theories during the first century of the field. Over the past three decades, however, progress and breakthroughs in quantum information has increasingly contributed to advances in nearly all research areas of fundamental physics \cite{Horodecki,terhal2003entanglement,calabrese2004entanglement,ryu2006,preskill2007,solodukhin2011review}. Characterizing the entanglement structure of many-body systems has become crucial for the development of new numerical techniques and quantum simulation methods \cite{laflorencie2016quantum,Eisert2010,osterloh2002scaling,vedral2008rmp}. More broadly, together with broken symmetry and topology, many-body entanglement is now recognized as one of the fundamental organizing principles of quantum matter \cite{wen2006,Kitaev2003,Kitaev2006,Haldane2008,Nussinov2009,Fidkowski2010,Fidkowski2011,jiang2012identifying}.

While the importance of entanglement in modern theories of quantum matter is undeniable, its impact on experiments has remained surprisingly limited. The standard framework for connecting many-body theory with experiment is linear response theory, epitomized by the Kubo formulas for various response functions. In the conventional textbook treatment, entanglement rarely appears in either the derivation or interpretation of these functions. A key reason why entanglement is difficult to characterize in many-body systems is precisely the absence of a straightforward connection between experimentally accessible linear responses and entanglement.

\begin{figure}[h]
    \centering
    \includegraphics[width=.99\columnwidth]{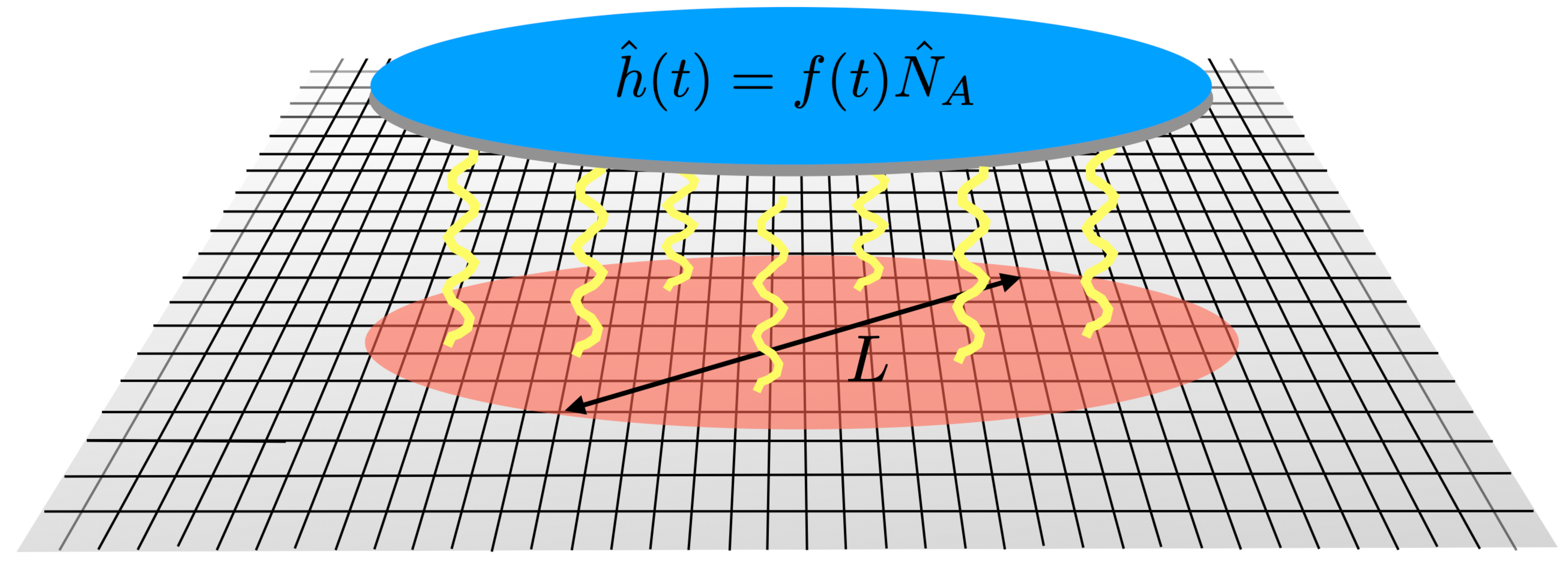}  
    \caption{Conserved lattice fermions exposed to a perturbation which couples to the particles on subsystem $A$ (red region) $\hat{N}_A=\sum_{i\in A}\hat{n}_i$ with linear dimension $L$. The leading order scaling in $L$ of various physical responses follow the same form as the scaling of the entanglement entropy of subsystem $A$.  }
    \label{fig:schematic}
\end{figure}

Detecting many-body entanglement, for instance via entanglement entropy, 
is now possible in engineered systems like cold atoms and trapped ions \cite{Greiner2015measuring,brydges2019probing}, and limited to small sizes due to exponential complexity of the quantum state tomography \cite{Cramer2010tomography,Eisert2010}. 
In bulk quantum matter, however, direct detection of entanglement remains even more challenging. 
Researchers have therefore turned to indirect signatures. 
One example is the quantum Fisher information (QFI), a quantum coherence estimator that measures how strongly an observable fails to commute with the state \cite{Smerzi_QFI_vs_entanglement,Zoller2016measuring_entanglement}. QFI can be inferred from nonlocal dynamical responses. A simpler proposal is the \emph{quantum variance}, defined as the difference between fluctuations and the isothermal susceptibility \cite{Roscilde2016,Roscilde2019}. 
Yet these connections between entanglement and fluctuations are usually model-specific and often rely on rather intricate and experimentally-elusive mathematical constructions such as QFI or full counting statistics \cite{KlichLevitov2009,LeHur2012,LeHur2012PRL}.

The purpose of this work is to establish a general connection between linear response functions and 
the scaling laws of entanglement entropy, providing strong constraints on the observable behavior of 
fermionic systems with conserved particles. This formulation can be viewed as an extension of the 
known relation between entanglement scaling and static fluctuations. 
Because the scaling of entanglement entropy is among the most robust signatures of quantum correlations 
in many-body systems, its appearance in linear response functions links two central but previously 
distinct concepts.
Specifically, we consider a Fermionic system exposed to perturbations which couple to the particles of a finite-size subsystem with linear dimension $L$, as illustrated in Fig.~\ref{fig:schematic}. Based on general considerations, we propose that the particle response and the energy-absorption rate in the ground state, both of which are traditionally expected to scale as the volume of the subsystem, should instead scale as the entanglement entropy of the subsystem. This scaling is illustrated for various free fermion states and shown to lead to several striking entanglement-limited responses.

Paper is organized so that in Sec.~\ref{sec:scaling laws}, we summarize the central properties of the entanglement entropy and its recently observed connections to static fluctuations. Then we explore the implications of the entanglement scaling to time-dependent fluctuations and various response functions. The section is concluded by discussing the physical consequences of entanglement-limited responses. In Sec.~\ref{sec:results}, we illustrate the postulated entanglement scaling properties for various free fermion states and provide a summary and outlook in Sec.~\ref{sec:Summary}.

\section{Entanglement scaling laws in linear response: general considerations}\label{sec:scaling laws}

In this section we review the scaling behavior of entanglement in many-body systems and the established connection between static fluctuations and the scaling of the von Neumann entropy. Building on these results, we extend the discussion to time-dependent fluctuations and linear-response functions, and highlight the unexpected physical implications of entanglement-limited response.

\subsection{Scaling of entanglement entropy and  static fluctuations }

We begin by summarizing the central properties of entanglement scaling laws in many-body systems. 
The von Neumann entanglement entropy of a pure state $\ket{\Psi}$ quantifies the bipartite entanglement 
between a subsystem $A$ and its complement $B$ as 
\[
\mathcal{S}_{vN} = -\mathrm{Tr}_{A}\, \rho_A \log \rho_A \equiv -\mathrm{Tr}_{B}\, \rho_B \log \rho_B,
\]
where $\rho_A = \mathrm{Tr}_B(\ket{\Psi}\bra{\Psi})$ and 
$\rho_B = \mathrm{Tr}_A(\ket{\Psi}\bra{\Psi})$ are the reduced density matrices of subsystems $A$ and $B$, respectively. 
Here we have used the fact that, in bipartite systems, the entanglement entropy does not depend on which subsystem is considered, so that 
$\mathcal{S}_{vN}[\rho_A] = \mathcal{S}_{vN}[\rho_B]$, and thus we dropped the subsystem dependence above.

The eigenstates of many-body systems with local interactions exhibit characteristic spatial scaling,
\begin{equation}\label{eq:scaling1}
\mathcal{S}_{vN} \propto F(L),
\end{equation}
where the subsystem-size dependence is captured by a universal function $F(L)$ that depends only on the type of state, 
the dimensionality of the subsystem, and its characteristic linear extent $L$. 
Typical leading-order behaviors are
\begin{equation}
\begin{array}{ll}
\text{Area law:}   & F(L) \sim L^{D-1}, \\
\text{Volume law:} & F(L) \sim L^{D}, \\
\text{Critical (gapless):}   & F(L) \sim L^{D-1}\log L .
\end{array}
\end{equation}
In general, gapped ground states are expected to follow the area law, 
while gapless ground states obey the critical scaling. 
States with finite energy density, on the other hand, exhibit volume-law scaling. 
A subsystem may also show volume-law scaling in the ground state 
if it has a lower effective dimensionality than the total system.

While the prefactor of the entanglement entropy can reveal important microscopic information about a many-body state, 
its most significant feature is usually its spatial scaling. 
Because the entanglement entropy is generally not directly measurable in experiments, 
it is desirable to access its scaling through more direct observables. 
Since entanglement scaling encodes strong constraints on quantum correlations, 
it is natural to ask how these constraints appear in standard experimental probes of many-body systems. 

Bipartite fluctuations of physical observables provide a natural candidate. 
In systems with particle-number conservation, the fluctuations of particle number between subsystems are particularly suitable. 
Here we consider a $D$-dimensional lattice of non-interacting fermions with a conserved total particle number $
\hat{N}=\sum_{i}\hat{n}_i$, meaning that $[\hat{H},\hat{N}]=0$. 
Here $\hat{H}$ is the Hamiltonian, $\hat{n}_i=\hat{c}^\dagger_i\hat{c}_i$ is the local particle-number operator on site $i$, 
and $\hat{c}^\dagger_i,\hat{c}_i$ are fermionic creation and annihilation operators. The indices for additional quantum numbers (spin, orbital, etc,) have been omitted for the clarity of expressions.
In a bipartition of the system into subsystems $A$ and $B$, we define the subsystem particle operators
\begin{equation}
\hat{N}_A=\sum_{i\in A}\hat{n}_i,
\end{equation}
and similarly for $B$. 

For systems with conserved total particle number, the particle-number fluctuations
\begin{align}
\delta^2 \hat{N}_{A} = \langle \hat{N}_{A}^2 \rangle - \langle \hat{N}_{A} \rangle^2
\end{align}
in an eigenstate of $\hat{H}$ share key features with the entanglement entropy. 
Since eigenstates are labeled by the total particle number $N$, the subsystem operators satisfy $\hat{N}_{A}=N-\hat{N}_{B}$. 
Hence, by analogy with $\mathcal{S}_{vN}[\rho_A] = \mathcal{S}_{vN}[\rho_B]$, the fluctuations obey
$\delta^2 \hat{N}_{A}=\delta^2 \hat{N}_{B}$. 
Moreover, $\delta^2 \hat{N}_{A}=\delta^2 \hat{N}_{B}=0$ if the Hamiltonian does not couple the subsystems. 
These properties motivate using fluctuations of conserved quantities as probes of entanglement. 

When the observable of interest is extensive, such as the subsystem particle number, these fluctuations not only signal entanglement but also share its spatial scaling,
\begin{equation}\label{eq:scaling2}
\delta^2 \hat{N}_{A}\propto F(L),
\end{equation}
where all subsystem-size dependence is contained in the universal function $F(L)$ 
appearing in the entanglement entropy scaling of Eq.~\eqref{eq:scaling1}. 
Because there is no general proof of Eq.~\eqref{eq:scaling1}, there is naturally no general proof of Eq.~\eqref{eq:scaling2}. 
Nevertheless, for the area-law case can be argued as follows.

The subsystem fluctuations can be rewritten as
\begin{align}
\delta^2 \hat{N}_{A} 
&= \langle  \hat{N}_{A}^2 \rangle - \langle  \hat{N}_{A} \rangle^2 \nonumber \\
&= \langle  \hat{N}_{A} (N -  \hat{N}_{B}) \rangle 
   - \langle  \hat{N}_{A} \rangle \left( N - \langle  \hat{N}_{B} \rangle \right) \nonumber \\
&= -\langle  \hat{N}_{A}  \hat{N}_{B} \rangle + \langle  \hat{N}_{A} \rangle \langle  \hat{N}_{B} \rangle 
   \equiv -\langle  \hat{N}_{A}  \hat{N}_{B} \rangle_c,
\end{align}
where $\langle \hat{N}_A\hat{N}_B \rangle_c
 = \langle \hat{N}_A\hat{N}_B \rangle - \langle \hat{N}_A \rangle \langle \hat{N}_B \rangle$
is the connected correlator. 
Expressing the subsystem operators through onsite operators gives
\begin{align}\label{eq:sumform1}
\delta^2 \hat{N}_{A}
= -\langle \hat{N}_{A} \hat{N}_{B} \rangle_c
= -\sum_{i \in A} \sum_{j \in B} \langle \hat{n}_{i} \hat{n}_{j} \rangle_c.
\end{align}
Thus the subsystem fluctuations are determined by local correlations between the two subsystems. In an area-law state the particle correlations are short-ranged,
\begin{align}\label{eq:damping1}
\langle  \hat{n}_{i} \hat{n}_{j} \rangle_c \sim e^{-|\vec{r}_i - \vec{r}_j| / \xi},
\end{align}
which implies that only the sites $i,j$ in the vicinity of the boundary layer (with a characteristic depth $\xi$) at the interface of subsystems contribute to the fluctuations \eqref{eq:sumform1}.
Consequently the fluctuations obey the area law
$\delta^2 \hat{N}_{A}\propto L^{D-1}$,
where $D$ is the spatial dimension of subsystem $A$ and $L$ its linear extent. 
This argument relies only on the conservation of $\hat{N}$ and the short range of correlations, 
without detailed assumptions about the Hamiltonian or the lattice. 

Although the reasoning above applies strictly to the area-law case, it suggests that the connection between entanglement scaling and fluctuations is broadly insensitive to microscopic details. To date, the scaling law of the form \eqref{eq:scaling2} for conserved $U(1)$ charge 
has been demonstrated in free-fermion systems \cite{poyhonen2021observing,moghaddam2022}, 
one-dimensional (1D) spin models (equivalent to interacting fermion chains) \cite{poyhonen2021observing,glodzik2025}, 
and monitored quantum circuits \cite{moghaddam2023,poyhonen2025}.

\subsection{Entanglement scaling laws in time-dependent fluctuations and linear response functions}

Here we show that, assuming that the static fluctuations obey the scaling \eqref{eq:scaling2}, then time-dependent fluctuations and linear-response functions also exhibit the same entanglement scaling. 
This observation has far-reaching physical consequences. First, we consider the subsystem particle noise, or \emph{dynamical structure factor} as \cite{Vignale2008quantum},
\begin{align}\label{eq:noise}
S_{N_A}(\omega)
&=\int dt\, e^{i\omega t}\, \langle \delta\hat{N}_A(t)\delta\hat{N}_A(0)\rangle \nonumber\\
&=\sum_{i,j} p_i
   \bigl|\langle\psi_i|\delta\hat{N}_A|\psi_j\rangle\bigr|^2
   2\pi \delta(\omega-E_j+E_i),
\end{align}
where $\delta\hat{N}_A(t)=\hat{N}_A(t)-\langle\hat{N}_A\rangle$ denotes the particle-number operator with its expectation value subtracted. 
The second line of Eq.~\eqref{eq:noise} is the Lehmann representation of the noise in terms of the eigenenergies $E_i$ and eigenstates $\ket{\psi_i}$ of the Hamiltonian $\hat{H}$. 
Although our main focus is the ground-state behavior, for which $p_0=1$ and $p_i=0$ for $i>0$, we keep the expressions general.

The static density fluctuations follow from the noise and satisfy the sum rule
\begin{align}\label{sumrule1}
\delta^2 \hat{N}_{A}
   = \int_{0}^{\infty} \frac{d\omega}{2\pi} S_{N_A}(\omega)
   \propto F(L).
\end{align}
Because $S_{N_A}(\omega)$ is positive for all $\omega$, each frequency component is expected to scale individually as
\begin{align}\label{eq:scaling3}
S_{N_A}(\omega) \propto F(L).
\end{align}
We demonstrate this property for several free-fermion systems and expect it to hold more generally. 
For any system obeying the static scaling \eqref{eq:scaling2}, the function $F(L)$ places an upper bound on the growth of $S_{N_A}(\omega)$ with subsystem size. 
For instance, in area-law systems the ground-state density noise can grow only as fast as the boundary of the subsystem.

When the subsystem particle fluctuations exhibit the scaling \eqref{eq:scaling3}, 
the same scaling is directly inherited by the retarded subsystem density–density response function \cite{Vignale2008quantum},
\begin{align}\label{susc}
\chi_{N_A}(\omega)
&=\sum_{i,j} p_i \:
   \bigl|\langle\psi_i  |\delta \hat{N}_A|\psi_j\rangle\bigr|^2 \nonumber\\
&\times \left[
   \frac{1}{E_i-E_j+\omega+i\eta}
 + \frac{1}{E_i-E_j-\omega-i\eta}\right],
\end{align}
where $\eta$ is a positive infinitesimal. 
The fluctuation–dissipation theorem relates the imaginary (dissipative) part of the response to the noise,
\begin{align}\label{FD}
S_{N_A}(\omega)
   = 2 \bigl[1+n_B(\omega)\bigr] \operatorname{Im}\chi_{N_A}(\omega),
\end{align}
with $n_B(\omega)$ the Bose–Einstein distribution function. 
At zero temperature this simplifies to $S_{N_A}(\omega)=2\,\operatorname{Im}\chi_{N_A}(\omega)$. 
Furthermore, the real and imaginary parts of the retarded response are connected through the Kramers–Kronig relation \cite{Vignale2008quantum,chaikin1995CMT},
\begin{align}\label{eq:KK}
\operatorname{Re}\chi_{N_A}(\omega)
   = \mathcal{P}\!\int_{-\infty}^\infty
     \frac{d\omega'}{\pi}
     \frac{\operatorname{Im}\chi_{N_A}(\omega')}{\omega'-\omega}.
\end{align}
Thus, when the noise satisfies the scaling \eqref{eq:scaling3}, 
Eqs.~\eqref{FD} and \eqref{eq:KK} imply that the retarded particle response must also follow the spatial scaling
\begin{align}\label{eq:scaling4}
\chi_{N_A}(\omega) \propto F(L),
\end{align}
where the subsystem-size dependence is determined by the universal function $F(L)$.

\subsection{Entanglement-limited physical responses}

Above we have explained how the entanglement entropy scaling laws for static subsystem fluctuations are inherited by the time-dependent fluctuations and subsystem response functions. Here we summarize how the entanglement scaling in these quantities give rise to surprising and counterintuitive physical effects. Consider a planar setup in Fig.~\ref{fig:schematic}, where a finite size probe couple to the particles underneath. This could describe, for example, electric potential induced by a capacitor plate coupling to charged particles on a 2D electron system. The Hamiltonian of the perturbed system becomes $\hat{H}'=\hat{H}+\hat{h}(t)$, where $\hat{H}$ is the Hamiltonian of the system in the absence of the perturbation and the second term $\hat{h}(t)=-f(t)\,\hat{N}_A$ describes a time-dependent potential. According to the linear response theory, the subsystem particle response is given by 
\begin{align}\label{eq:response}
\langle \delta \hat{N}_{A}(\omega)\rangle= \chi_{N_A}(\omega)f(\omega),
\end{align}
where $\langle \delta \hat{N}_{A}(\omega)\rangle=\int dt\langle \delta \hat{N}_{A}(t)\rangle e^{-i\omega t}$. 
Now, the instantaneous power delivered by the driving or equivalently the total energy absorption rate is given by \cite{chaikin1995CMT},
\begin{align}
\dot{E}(t)
&=\frac{d}{dt}\langle \hat{h}(t)\rangle
   \approx \dot f(t)\,\delta\langle \hat N_A (t)\rangle \\
&=\!\int\!\frac{d\omega}{2\pi}
     \int\!\frac{d\omega'}{2\pi}
     (-i\omega)\,f(\omega)\,
     \chi(\omega')\,f(\omega')\,
     e^{-i(\omega+\omega')t} . \nonumber
\end{align}
Its long–time average is
\begin{align}
\overline{\dot{E}}
&=\lim_{T\to\infty}\frac{1}{T}\int_{0}^{T}dt\,\dot{E}(t) \nonumber\\
&=\int\frac{d\omega}{2\pi}
    (-i\omega)\,f(-\omega)\,
    \chi_{N_A}(\omega)\,f(\omega).
\end{align}
For a real external drive $f(t)$ one has $f(-\omega)=f(\omega)^{*}$. 
Since the average power must be real, this gives
\begin{align}
\overline{\dot{E}}
&=\int\frac{d\omega}{2\pi}\,
   \omega\,|f(\omega)|^2\,
   \operatorname{Im}\chi_{N_A}(\omega) \nonumber\\
&=\int\frac{d\omega}{4\pi}\,
   \omega\,|f(\omega)|^2\,
   S_{N_A}(\omega), \label{eq:absorption}
\end{align}
where, in the second line, we used the zero-temperature
relation $\operatorname{Im}\chi_{N_A}(\omega) = S_{N_A}(\omega)/2$.

When the scaling relations \eqref{eq:scaling2} and \eqref{eq:scaling3} hold, both the particle response \eqref{eq:sumform1} and the absorption power \eqref{eq:absorption} inherit the subsystem-size scaling dictated by the entanglement–entropy function $F(L)$. This behavior marks a striking departure from the conventional expectation that these quantities should scale extensively with the perturbed volume. For instance, in a 2D insulator obeying an area-law entanglement scaling, the particle response and the absorption rate scale with the \emph{boundary length} of the capacitor plate, rather than with the total perturbed area beneath the plate. Moreover, this conclusion is not confined to low frequencies but holds quite generally. In metallic systems, where the ground-state entanglement is also subextensive, the ground-state response to a perturbation coupling to conserved particles is likewise subextensive. This provides a clear example of how entanglement scaling can strongly influence the physical ground-state response and how such effects can be directly probed through measurable physical responses.


\section{Entanglement-limited scaling of the dynamical structure factor for free fermions}
\label{sec:results}

In the previous section, we argued that, in particle-conserving systems, 
the dynamical structure factor obeys the same spatial scaling 
\eqref{eq:scaling3} as the entanglement entropy \eqref{eq:scaling1}. 
We also showed that the linear-response quantities 
\eqref{eq:response} and \eqref{eq:absorption} inherit this entanglement-driven scaling. 
Here we explicitly demonstrate the scaling relation \eqref{eq:scaling3} 
for a variety of free-fermion models. We consider noninteracting fermions on a regular lattice with periodic 
boundary conditions, described by the Hamiltonian
\begin{align}
H = \sum_{\vec{k},\sigma ,\sigma'} 
c_{\vec{k}\sigma'}^\dagger \,
H_\vec{k}^{\sigma'\sigma} \,
c_{\vec{k}\sigma},
\end{align}
where $\vec{k}$ is the $D$-dimensional  crystal momentum and 
$\sigma,\sigma'$ label orbital indices.  
Such free-fermion systems can be diagonalized in a Bloch basis 
$\lvert \omega_{\vec{k}m} \rangle$, related to the momentum operators by  
\(
c_{\vec{k}\sigma}
= \sum_m \langle \sigma \vert \omega_{\vec{k}m} \rangle d_{\vec{k}m}.
\)  
The Hamiltonian then takes the diagonal form
\begin{align}
H = \sum_{\vec{k},m} \omega_{\vec{k}m}\,
d_{\vec{k}m}^\dagger d_{\vec{k}m},
\end{align}
where $m$ is the band index.

The dynamical structure factor \eqref{eq:noise} for such free-lattice models 
is derived in Appendix~\ref{app:a}.  
Because the exact spectrum is a sum of $\delta$-functions, 
it is convenient to introduce a frequency-smoothed (coarse-grained) 
structure factor,
\begin{align}
\overline S_{N_A}(\omega_0)
&=\int_{\omega_0-\Delta\omega}^{\omega_0+\Delta\omega} \frac{ d\omega}{2\pi}\: 
S_{N_A}(\omega) \\
&=\frac{1}{V^2}\sum_{\vec{j},\vec{j}'\in A}
\sum_{\vec{k}m}\sum_{\vec{k}'m'}
e^{i(\vec{k}-\vec{k}')\cdot(\vec{j}-\vec{j}')}
\mathrm{Tr}\!\left[P_{\vec{k}m} P_{\vec{k}'m'}\right] \nonumber\\
&\quad\times n_{\vec{k}m}
\left(1-n_{\vec{k}'m'}\right)
K\!\left(\omega_0+\omega_{\vec{k}m}-\omega_{\vec{k}'m'}\right), \nonumber
\end{align}
where $n_{\vec{k}m} \equiv n(\omega_{\vec{k}m}) =
\bigl[1 + e^{\beta \omega_{\vec{k}m}}\bigr]^{-1}$ 
is the Fermi-Dirac distribution at inverse temperature 
$\beta = 1/(k_B T)$, and $P_{\mathbf{k}m}=|\omega_{\mathbf{k}m}\rangle \langle \omega_{\mathbf{k}m}|$ is the projector onto the energy band $m$. 
The trace is taken over orbital indices. In the coarse-grained structure factor, the energy $\delta$-functions 
are replaced by a unit-box window given by the function 
\begin{align}
    K(\omega)=\frac{1}{2\Delta\omega}\Theta(\Delta\omega - |\omega | ),
\end{align}
where $\Theta$ represents the Heaviside step function.
The window $\Delta\omega$ is chosen significantly larger than the typical 
level spacing of the finite system, but much smaller than the 
characteristic frequency scale over which the structure factor varies.  
Within this regime, the functional form of the structure factor is largely 
insensitive to the details of the coarse graining, aside from small 
band-edge shifts of order $\Delta\omega$. Although we introduce coarse graining primarily for illustration purposes, the smoothed structure factor is also the physically 
relevant quantity in experiments where the frequency resolution 
cannot resolve individual energy levels. In what follows, we focus on ground-state properties at zero temperature.

\subsection{Area-law regime}
As we discussed before, ground states of systems with a finite energy gap to excited states are generally expected to obey the  area-law entanglement. A simple and well-studied class of such systems is provided by two-band insulators,  
described by the Bloch Hamiltonian
\begin{align}
H_{\vec{k}} = \vec{d}(\vec{k})\cdot \boldsymbol{\sigma},
\end{align}
where $\vec{d}(\vec{k}) = \bigl[d_x(\vec{k}), d_y(\vec{k}), d_z(\vec{k})\bigr]$ 
and $\boldsymbol{\sigma} = (\sigma_x,\sigma_y,\sigma_z)$ is the vector of Pauli matrices.  
The corresponding energy bands are
$\omega_{\vec{k}\pm} = \pm \bigl|\vec{d}(\vec{k})\bigr|$, with band projectors
\begin{align}
P_{\vec{k}\pm} = \tfrac{1}{2}\!\left[ \mathbbmss{1} \pm 
\widehat{\vec{d}}(\vec{k})\!\cdot\!\boldsymbol{\sigma} \right],
\end{align}
where $\widehat{\vec{d}}(\vec{k}) = \vec{d}(\vec{k})/|\vec{d}(\vec{k})|$, and $\mathbbmss{1}$ 
represents the $2\times 2$ identity matrix. At zero temperature, and with the Fermi energy lying inside the band gap 
(i.e., for the undoped model), the valence band is completely filled while the 
conduction band remains empty.  
In this case the coarse-grained dynamical structure factor takes the form
\begin{align} \label{eq:noise2}
\overline S(\omega_0)
&= \frac{1}{2V^2} \sum_{\vec{j},\vec{j}' \in A} 
   \sum_{\vec{k},\vec{k}'}
   e^{i(\vec{k}-\vec{k}')\cdot(\vec{j}-\vec{j}')}
   \left[1 - \widehat{\vec{d}}(\vec{k})\!\cdot\!\widehat{\vec{d}}(\vec{k}')\right] \nonumber\\
&\quad\times K\bigl(\omega_0 + \omega_{\vec{k}-} - \omega_{\vec{k}'+}\bigr),
\end{align}
where $K(\omega)$ denotes the frequency-smoothing kernel introduced earlier.

\begin{figure}[t!]
\centerline{\includegraphics[width=0.96\linewidth]{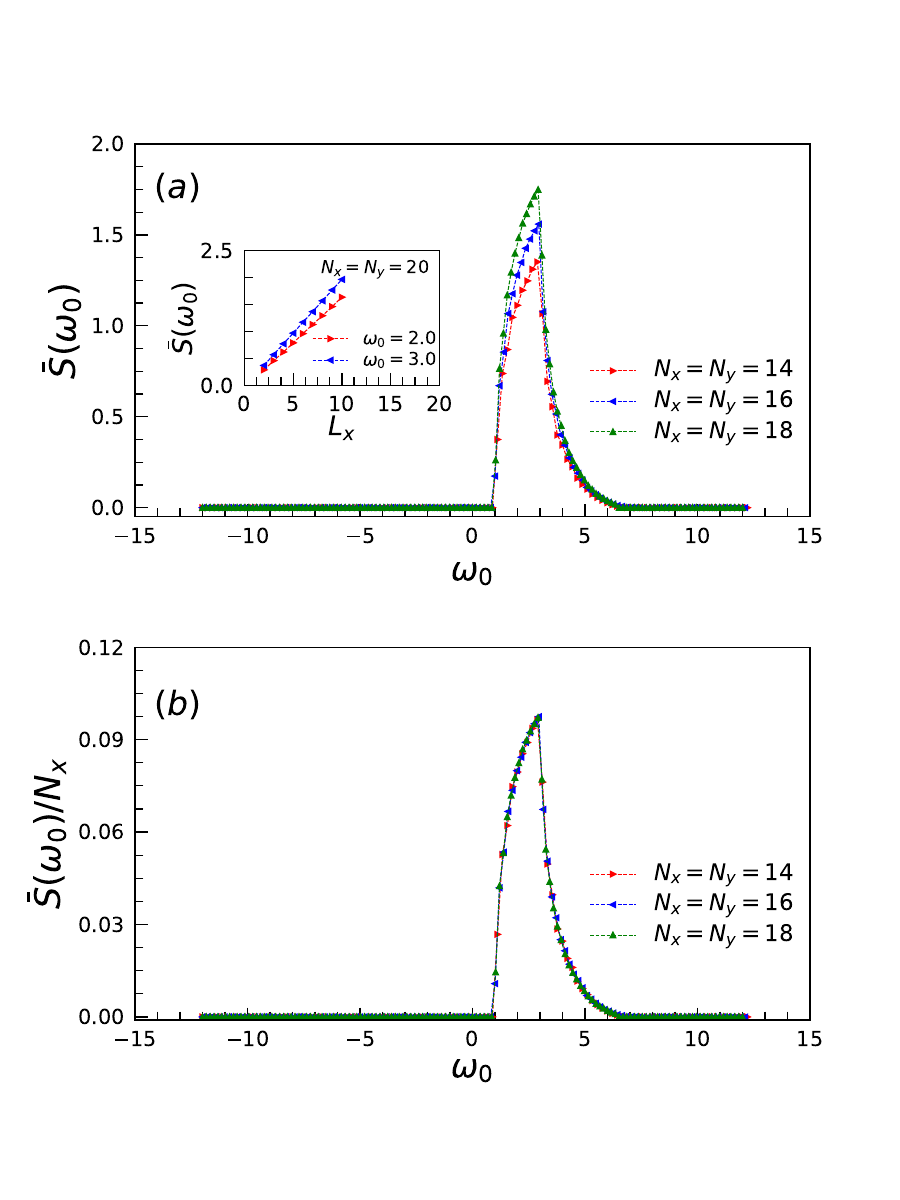}}
\caption{(Color online) 
Behavior of the dynamical noise in the QWZ model with $m=1.0$.  
For all system sizes the subsystem is chosen as $l_x = N_x/2$ and $l_y = N_y/2$.  
The coarse-graining width is fixed for $N_x = N_y = 12$ as 
$\Delta \omega_0 = 2 \delta \omega $, where
$
\delta \omega =
\max_{\bf k}\big|
\omega_{{\bf k}+\delta {\bf k}} -
\omega_{\bf k}
\big|
$
with ${\bf k}$ are discrete momentum values (due to finite system lengths) in a mesh grid given by Eq. \eqref{eq:discrete-k-2D},
and momentum step $\delta {\bf k} = (2\pi/N_y)\hat{\bf y}$.
Since, for the set of parameters considered here, the valence and conduction bands span 
$\omega_{\pm} \in (\pm 1, \pm 3)$, the dynamical structure factor is ideally nonzero only 
for $\omega_0 \in (2,6)$, corresponding to possible transitions from valence to conduction 
states, when the frequency window $\Delta \omega$ is very small. For a finite $\Delta \omega$, 
however, we observe a slight broadening of the range over which $\bar{S}(\omega_0)$ remains finite.
}
\label{Fig:2D2B}
\end{figure}

As a prototypical model, we consider Qi-Wu-Zhang (QWZ) Hamiltonian 
\begin{align}\label{eq:QWZ}
H_\vec{k} = m_{\bf k}\:{\sigma _z} + \sin {k_x}\:{\sigma _x} + \sin {k_y}\: {\sigma _y},    
\end{align}
where $m_{\bf k} = m - \cos {k_x} - \cos {k_y}$ in which the parameter $m$ controls the size of the bulk energy gap.  
We evaluate the dynamical structure factor for rectangular subsystems of various sizes with the results shown in Fig.~\ref{Fig:2D2B}. For the system sizes considered in our calculations \cite{footnote:computation}, the structure factor already exhibits a remarkably precise area-law scaling across a wide range of frequencies as it can be seen from Fig. \ref{Fig:2D2B}(a) where $\overline S(\omega_0)$ is shown as function of central frequency $\omega_0$ for different subsystem sizes. The linear variation with length, corresponding to are-law behavior is more clearly shown in the inset for two different frequencies. 
Moreover, the curves of $\overline S(\omega_0)$ for different subsystem sizes collapse to a single curve when scaled by the  linear dimension of the subsystem, as illustrated in \ref{Fig:2D2B}(b). 

To obtain the results shown here, we use a coarse–graining window
$\Delta \omega_0 = 2 \delta \omega $, set for the reference size $N_x = N_y = 12$.  
Here $\delta \omega $ denotes the largest single–step energy spacing in the discrete momentum mesh,
$\delta \omega 
= \max_{\mathbf{k}}
\big|
\omega_{\mathbf{k}+\delta \mathbf{k}} -
\omega_{\mathbf{k}}
\big|$,
where the allowed discrete momenta ${\bf k}\equiv \mathbf{k}_{j_x,j_y}$ are
\begin{align} \label{eq:discrete-k-2D}
\mathbf{k}_{j_x,j_y} =
\left(-\pi + \frac{2\pi j_x}{N_x}  \right)\hat{\mathbf{x}}
+
\left(-\pi + \frac{2\pi j_y}{N_y} \right)\hat{\mathbf{y}},
\end{align}
with $j_x = 1,\dots,N_x$ and $j_y = 1,\dots,N_y$, and the momentum step in the
$y$–direction is
$\delta \mathbf{k} = (2\pi/N_y)\hat{\mathbf{y}}$.

\subsection{Critical gapless (metallic) phase}

\begin{figure}[t]  
\centerline{\includegraphics[width=0.96\linewidth]{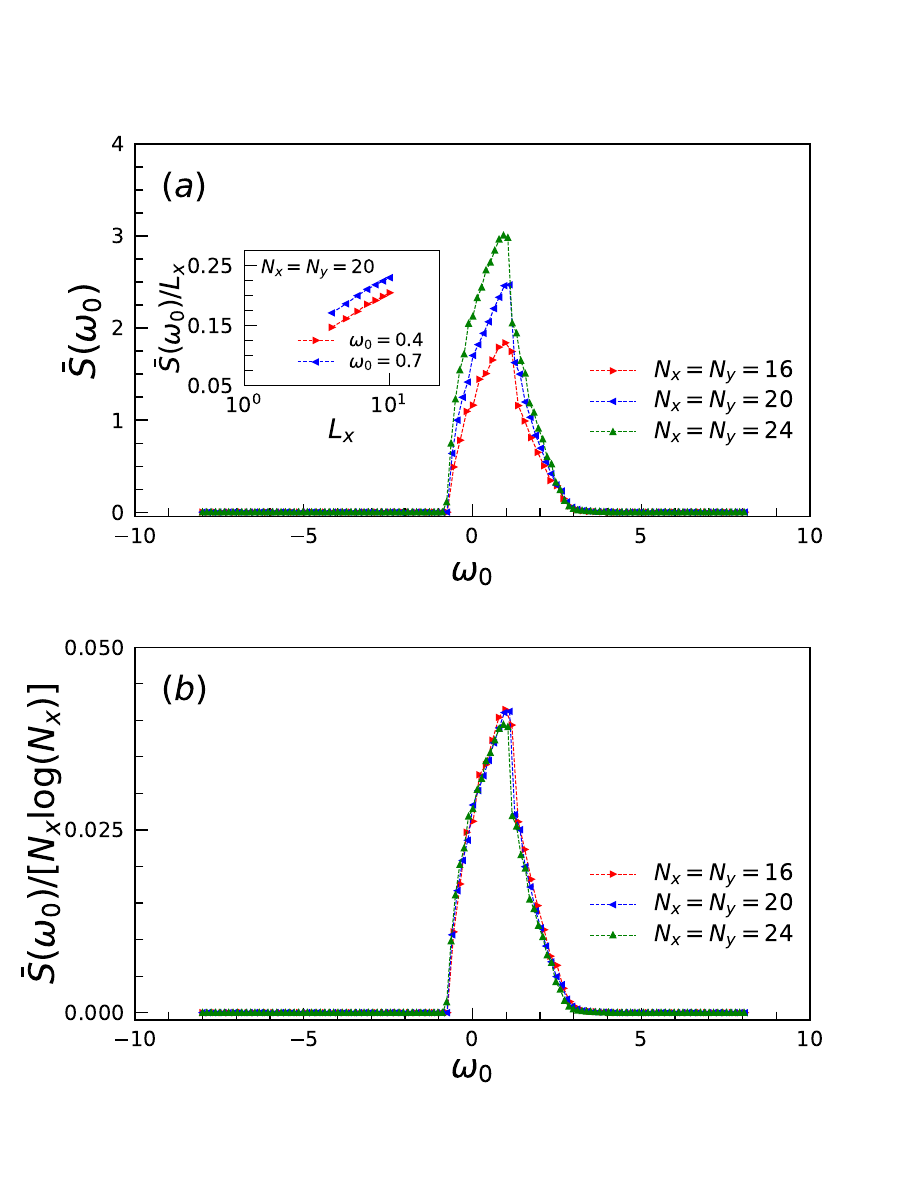}}
\caption{(color online) Noise power in a 2D metallic lattice model
with single–band dispersion
$\omega_{\mathbf{k}} =
- t \bigl[\cos(k_x) + \cos(k_y)\bigr] + \mu$,
where $t = 1.0$ and $\mu = 0.5$.
For all system sizes the subsystem is chosen as
$l_x = N_x/2$ and $l_y = N_y/2$.
The frequency-smoothing width is fixed for
$N_x = N_y = 14$ as
$\Delta \omega_0 = 2 \delta \omega $, where
$\delta \omega 
= \max_{\mathbf{k}}
\big|
\omega_{\mathbf{k}+\delta \mathbf{k}} -
\omega_{\mathbf{k}}
\big|$.
The discrete momentum values $\mathbf{k}$ are defined on the same mesh
introduced in Eq.~\eqref{eq:discrete-k-2D}, and the momentum step
in the $y$–direction is
$\delta \mathbf{k} = (2\pi/N_y)\hat{\mathbf{y}}$. 
} 
\label{Fig-2D1B}
\end{figure}

{An interesting case arises in critical gapless systems as well as metallic phases, which exhibit the characteristic critical-entropy scaling of entanglement 
$\propto l^{D-1}\log l$ with subsystems length scale $l$, 
where $D$ denotes the spatial dimension~\cite{Wolf2006,Swingle2010,Klich2006Widom}. When subsystem charge fluctuations are used as a proxy for entanglement, these systems are also known to display a similar logarithmic enhancement of the area-law behavior \cite{LeHur2012PRL,poyhonen2021observing,moghaddam2022}.}

\begin{figure}[t]
\centerline{\includegraphics[width=0.96\linewidth]{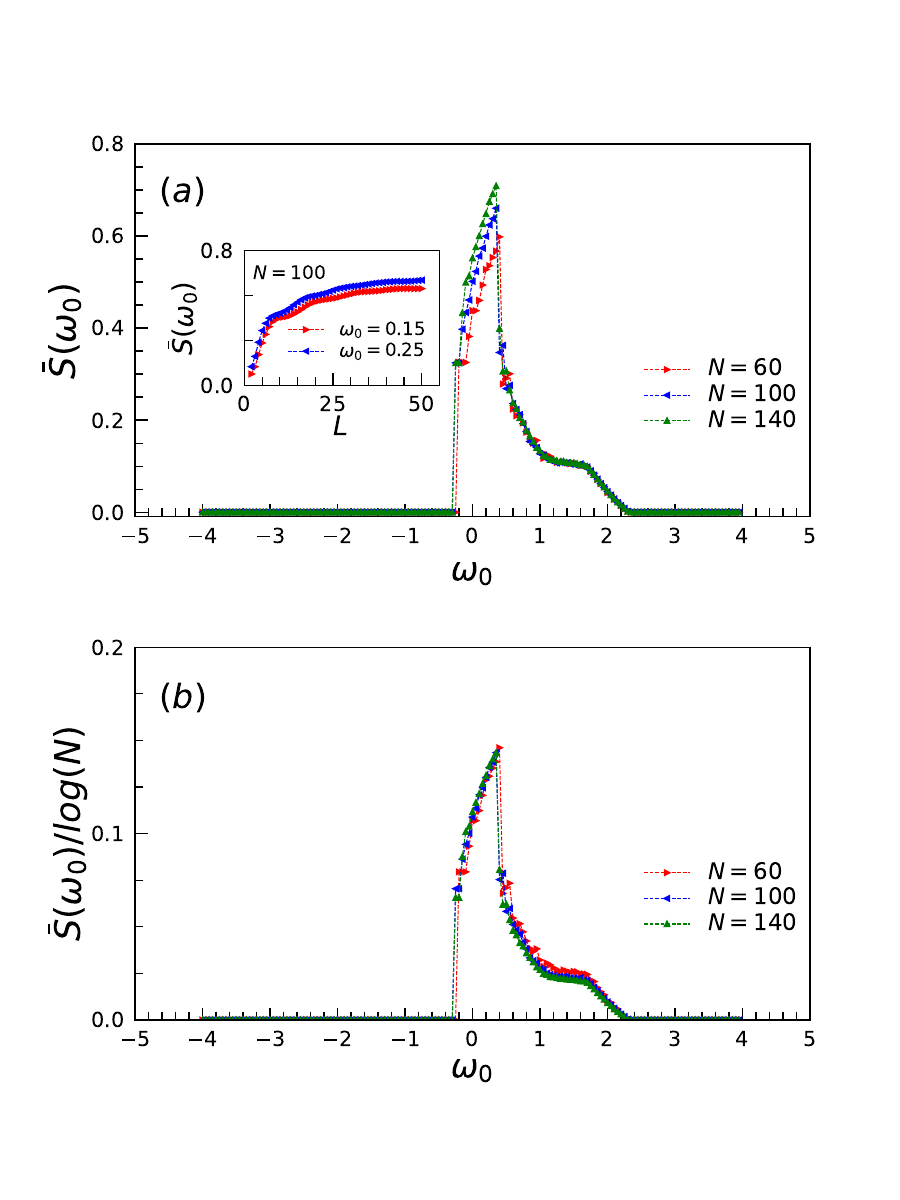}}
\caption{(color online) Behavior of the noise power in a 1D one-band model. 
The dispersion relation is given by $\omega_{k} = -t\cos(k) + \mu$, where we use the parameters 
$t = 1.0$ and $\mu = 0.5$. The subsystem length is fixed at $l = N/2$. 
The frequency-swapping window $\Delta\omega_0 = 2\delta\omega$ is obtained following the same 
procedure as in Fig.~\ref{Fig-2D1B}, but applied to the 1D case and for a fixed system size $N = 40$.
}
\label{Fig-1D1B}
\end{figure}

To explore the behavior of the dynamical structure factor (noise power) in this regime, we consider a single-band metal described by a simple tight-binding Hamiltonian
\begin{align}\label{eq:2d_metal}
    H_{k} = -t_1 \cos(k_x) - t_2 \cos(k_y) + \mu,
\end{align}
with dispersion relation $\omega_{\mathbf{k}} \equiv H_{\mathbf{k}}$. At zero temperature, the noise power reduces to
\begin{align}
\overline S (\omega_0) = \frac{1}{V^2} \sum\limits_{\vec{j},\vec{j}' \in A} 
\sum\limits_{\vec{k},\vec{k}'} 
e^{i(\vec{k}-\vec{k}')\cdot(\vec{j}-\vec{j}')} 
K(\omega_0 + \omega_{\vec{k}} - \omega_{\vec{k}'}),
\end{align}
and the corresponding results for a square-shaped subsystem with $l = N_x/2 = N_y/2$ and fixed chemical potential $\mu = 0.5$ and hoppings $t_1 = t_2 = 1$ are shown in Fig.~\ref{Fig-2D1B}(a) for three different system sizes. To demonstrate that the noise power exhibits an $l \log l$ dependence on the subsystem length $l$, we scale the noise power by $N_x \log N_x$ in Fig.~\ref{Fig-2D1B}(b), where an almost perfect collapse of the data onto a single curve is observed across all frequencies. As an additional quantitative check of the $l \log l$ behavior, the inset of Fig.~\ref{Fig-2D1B}(a) shows the scaled noise power $\bar{S}(\omega_0)/l$ as a function of the subsystem length $l$ in logarithmic scale. For the hopping parameters considered, the model has a bandwidth of $4$, implying that the dynamical structure factor is ideally nonzero only for $\omega_0 \in (0,4)$, corresponding to allowed transitions from occupied to unoccupied states when the frequency window $\Delta \omega$ is very small. For a finite $\Delta \omega$, however, a slight broadening of the frequency range over which $\bar{S}(\omega_0)$ remains finite is observed.

We now consider the corresponding metallic behavior in a 
1D one-band model. As shown in Fig.~\ref{Fig-1D1B}, the noise power for the 
dispersion $\omega_k = -t\cos(k) + \mu$ with $t = 1$ and $\mu = 0.5$ exhibits the expected 
logarithmic scaling characteristic of 1D gapless metallic systems. 
Fixing the subsystem size to $l = N/2$, we observe that $\overline{S}(\omega_0)$ grows as 
$\log N$. The logarithmic behavior is verified both by considering the scaled noise power 
$\bar{S}(\omega_0)/\log l$ in Fig.~\ref{Fig-1D1B}(b), and by plotting $\bar{S}(\omega_0)$ as a 
function of subsystem length in a semi-logarithmic scale, shown in the inset of 
Fig.~\ref{Fig-1D1B}(a). 

\begin{figure}[t]
\centerline{\includegraphics[width=0.96\linewidth]{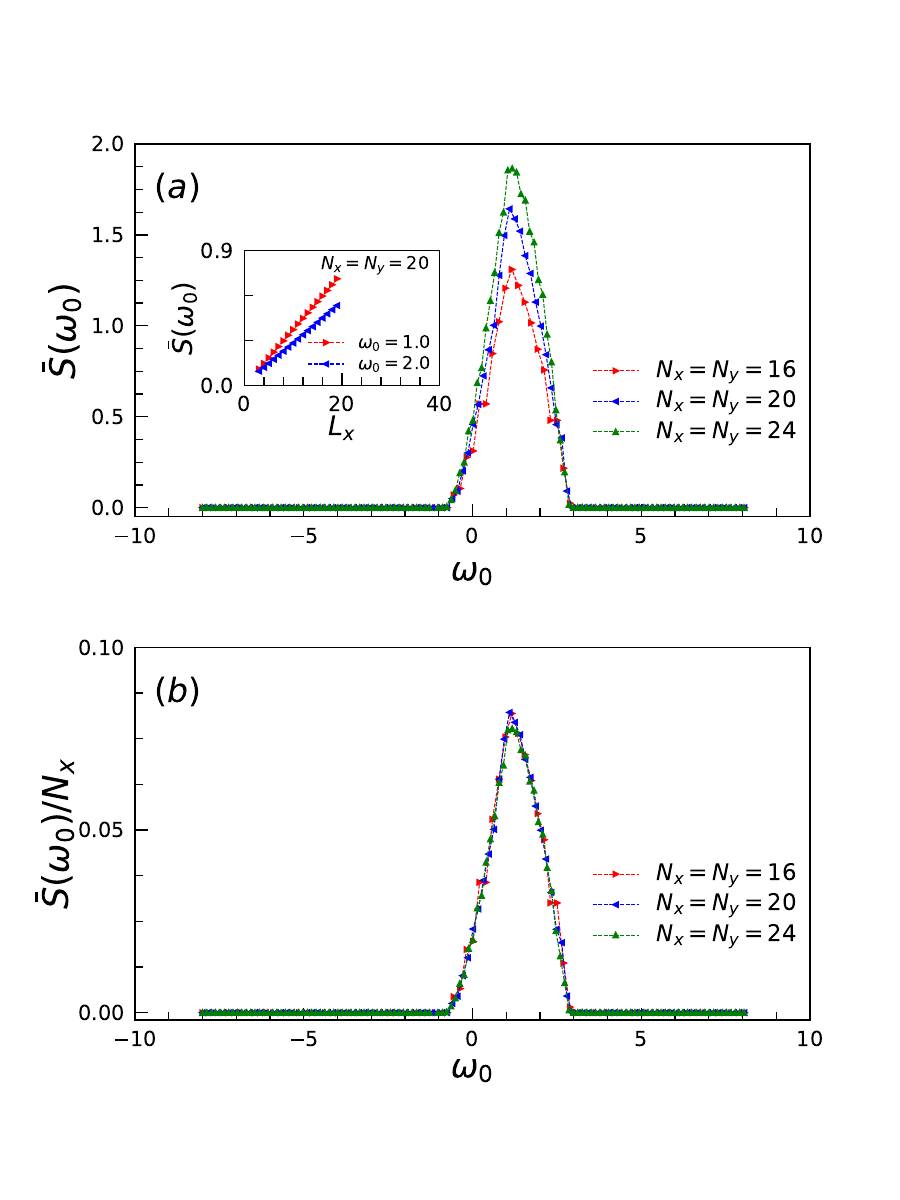}}
\caption{(color online) Behavior of the noise power in a 1D subsystem 
($l_x = N_x$ and $l_y = 1$) of a two-dimensional single-band model on a square lattice. 
The frequency window $\Delta \omega_0 = 2\delta\omega$ is fixed based on the value of 
$\delta\omega$ obtained for a system of size $N_x = N_y = 14$ ($\delta\omega$ is obtained in the same way as for the results shown in Fig.~\ref{Fig-2D1B}). 
Panel (a) shows the unscaled noise power, while panel (b) presents the same data scaled 
as $\bar{S}(\omega_0)/N_x$. The 1D subsystem embedded in the 2D system clearly exhibits a 
one-dimensional volume-law behavior, evidenced by the collapse of all curves in the scaled 
results of panel (b), as well as the linear dependence on the subsystem length $l_x$ shown in the inset of panel (a).
}
\label{Fig-2D1B-PRX}
\end{figure}

The frequency-swapping window $\Delta\omega_0 = 2\delta\omega$, computed using the same procedure 
as for the 2D case, produces a smooth dependence of the noise power on $\omega_0$, 
with the nonzero support bounded by the 1D bandwidth. Moreover, for the hopping 
parameters considered, the bandwidth of the model is $2$, which determines the approximate 
frequency range over which nonzero noise power is expected (the small deviations arise from the 
finite broadening used in the frequency-swapping procedure via $\Delta\omega_0$). The results of this subsec. confirm 
the dimensional dependence of charge-fluctuation scaling in a frequency resolved form and demonstrate that the dynamical 
structure factor faithfully captures the characteristic $\bar{S}(\omega_0)\propto l^{D-1}\log l$ behavior for two- and 1D Fermi systems ($D=1,2$).

\begin{figure}[t]
\centerline{\includegraphics[width=0.96\linewidth]{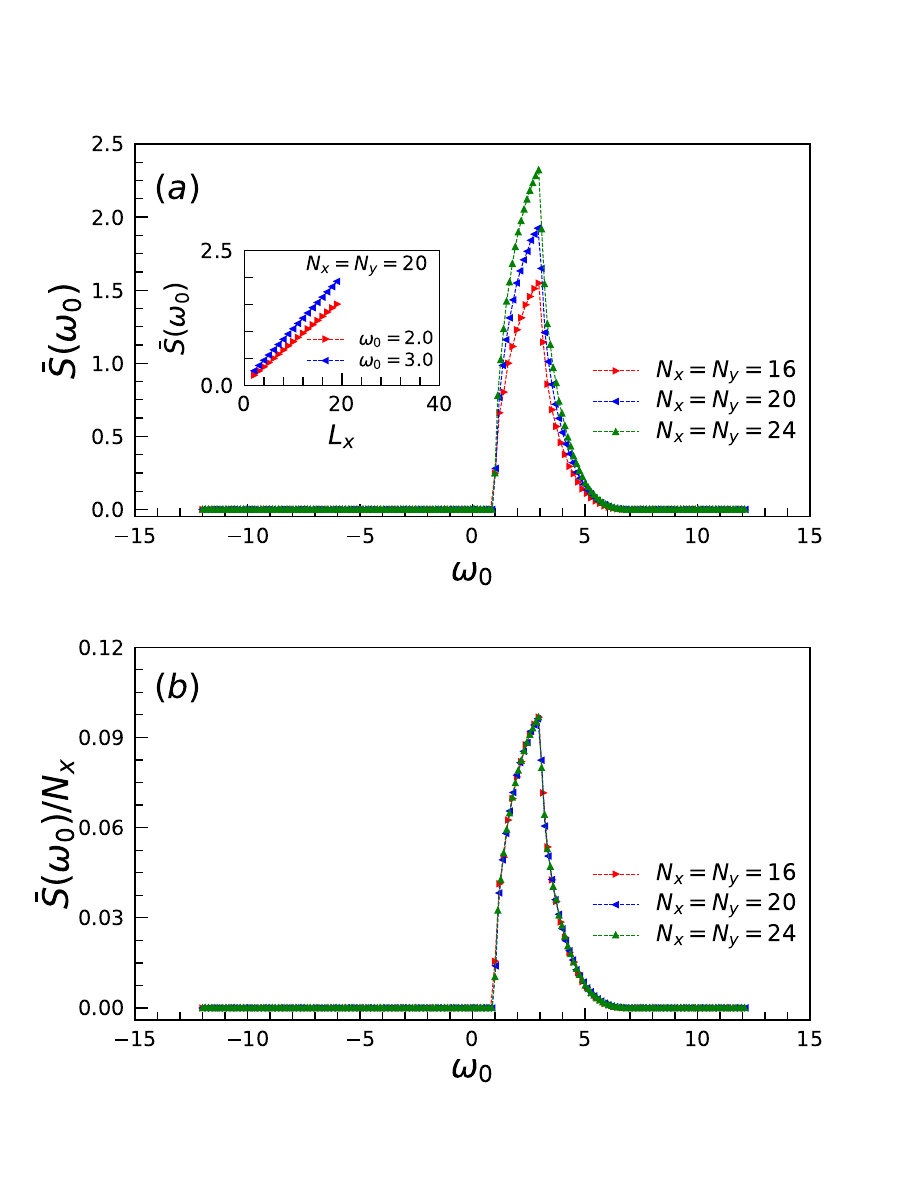}}
\caption{(color online) Noise power in a 1D subsystem of the two-dimensional QWZ lattice model. 
All parameters and procedures are identical to those used in Fig.~\ref{Fig-2D1B-PRX}, but applied here to the two-band QWZ model instead of the single-band metallic system.
}
\label{Fig-2D2B-PRX}
\end{figure}

\subsection{Volume–law regime in lower–dimensional subsystems}
While highly excited states with finite energy density typically exhibit volume-law entanglement scaling, the ground states of local Hamiltonians generally do not.  Recent work, however, has shown that the reduced density matrices of lower-dimensional subsystems embedded in a $D$-dimensional gapped Dirac-fermion vacuum behave similarly to thermal states with an effective finite temperature induced by ground-state entanglement \cite{moghaddam2022}. Consequently, for these lower–dimensional subsystems the entanglement entropy scales proportionally to the subsystem volume, even though the full $D$-dimensional ground state obeys the usual area law.

To explore how this emerged volume–law behavior is reflected in dynamical observables, we investigate the dynamical structure factor of 1D subsystems embedded in 2D parent systems. Specifically, we study both a metallic state and a gapped topological insulator described, respectively, by the Hamiltonians \eqref{eq:2d_metal} and \eqref{eq:QWZ}. This setting provides an additional and qualitatively distinct test of the connection between entanglement entropy and linear–response quantities. As illustrated in Fig.~\ref{Fig-2D1B-PRX} for the 2D metallic parent state and in Fig.~\ref{Fig-2D2B-PRX} for the QWZ model, the 1D subsystems in both cases display a clear volume-law scaling of the dynamical structure factor. This scaling closely tracks the subsystem entanglement entropy, demonstrating that the correspondence between entanglement scaling and 
linear–response functions persists even when the parent states themselves obey different entanglement laws.


\section{Summary and outlook}\label{sec:Summary}

In this work, we proposed that for particle-conserving fermionic systems, ground-state particle 
fluctuations within a finite region, together with the associated linear response functions, 
exhibit the same spatial scaling as the entanglement entropy. We explicitly demonstrated this 
scaling in free-fermion models and, building on earlier observations for static fluctuations, 
argued that the result can persist in interacting systems. Remarkably, and in contrast to 
conventional expectations, several linear-response quantities become subextensive at zero 
temperature. Most notably, the response and energy absorption rate in gapped phases (which obey 
an entanglement-entropy area law) scale with the boundary of the perturbed region.

Real physical systems, however, are never strictly in their ground state, raising the question of 
how entanglement-related scaling manifests in finite-temperature response functions. In agreement 
with standard expectations, finite-temperature responses scale extensively at leading order. 
Nevertheless, this does not preclude the experimental observation of ground-state scaling. This is 
particularly easier to achieve in the case of a gapped system: the structure factor 
\eqref{eq:noise} at finite temperature takes the form 
$S_{N_A}(\omega) = a(T) L^{D-1} + b(T) L^D$, where the volume-law coefficient satisfies 
$b(T) \to 0$ as the temperature $T$ approaches zero. Consequently, there exists a 
temperature-dependent crossover length scale $L_c(T) = a(T)/b(T)$ below which area-law scaling 
dominates. In this way, entanglement scaling remains experimentally accessible even at finite but low enough
temperatures. While the precise value of $L_c(T)$ is model-dependent, for gapped systems the 
coefficient $b(T)$ is exponentially suppressed as $b(T) \sim e^{-\Delta/T}$ for all temperatures well below 
the gap $\Delta$. As a result, $L_c(T)$ becomes exponentially large at low temperatures, which 
favors the observation of the area-law regime. A detailed analysis of finite-temperature effects 
and the associated crossover behavior will be explored in future works.

\acknowledgements
 A.G.M. and T.O. acknowledge Jane and Aatos Erkko Foundation for financial support. T.O. also acknowledges the Finnish Research Council project 362573.

\appendix
\begin{widetext}
\section{Derivation of the dynamic structure factor for free fermion}
\label{app:a}
Let’s consider a free fermion system on a lattice with periodic boundary conditions with the generic Hamiltonian as
\begin{align}
H = \sum_{\vec{k},\sigma ,\sigma'} 
c_{\vec{k}\sigma'}^\dagger \,
H_\vec{k}^{\sigma'\sigma} \,
c_{\vec{k}\sigma} 
= \sum\limits_{\vec{k},m} \omega _{\vec{k} m}d_{\vec{k} m}^\dag  d_{km}
\end{align}
where $\vec{k}$ is an arbitrary dimensional crystal momentum, $\sigma$ accounts for internal degrees of freedom (orbital, spin, and sublattice indices) and $m$ is the energy band index in the diagonal basis. Considering a subsystem $A$ and the particle number operator $n = \sum\nolimits_{(\vec{j} \in A),\sigma } c_{\vec{j}\sigma }^\dag  c_{\vec{j}\sigma }$ where the sum is over the subsystem lattice sites $\vec{j}$ and orbitals $\sigma$, any external perturbation that couples to $n$ can be captured by a linear response function. By defining the particle number fluctuation operator as $\delta n = n - \langle n \rangle$, its temporal correlation would read
\begin{eqnarray}
{S_{\delta n}}(t) &=& \left\langle {\delta n(t)\delta n(0)} \right\rangle  = \left\langle {\left[ {n(t) - \left\langle {n(t)} \right\rangle } \right]\left[ {n(0) - \left\langle {n(0)} \right\rangle } \right]} \right\rangle \nonumber\\
 &=& \left\langle {n(t)n(0)} \right\rangle  - \left\langle {n(t)} \right\rangle \left\langle {n(0)} \right\rangle  \nonumber\\
 &=& \sum\limits_{(\vec{j},\vec{j}' \in A),\sigma ,\sigma '} {\left[ {\left\langle {c_{\vec{j}\sigma }^\dag (t){c_{\vec{j}\sigma }}(t)c_{\vec{j}'\sigma '}^\dag (0){c_{\vec{j}'\sigma '}}(0)} \right\rangle  - \left\langle {c_{\vec{j}\sigma }^\dag (t){c_{\vec{j}\sigma }}(t)} \right\rangle \left\langle {c_{\vec{j}'\sigma '}^\dag (0){c_{\vec{j}'\sigma '}}(0)} \right\rangle } \right]} \nonumber\\
 &=& \sum\limits_{(\vec{j},\vec{j}' \in A),\sigma ,\sigma '} {\left\langle {c_{\vec{j}\sigma }^\dag (t){c_{\vec{j}'\sigma '}}(0)} \right\rangle \left\langle {{c_{\vec{j}\sigma }}(t)c_{\vec{j}'\sigma '}^\dag (0)} \right\rangle } 
\end{eqnarray}
that we have used Wick's theorem and the fact that for normal metals or insulators we have $\langle {c_{\vec{j}\sigma }^\dag (t)c_{\vec{j}'\sigma '}^\dag (0)}\rangle =\langle {{c_{\vec{j}\sigma }}(t){c_{\vec{j}'\sigma '}}(0)}\rangle =0$ \footnote{These terms exist in fermionic systems with superconducting correlations}. Now applying Fourier transformation $c_{\vec{j}\sigma } = \frac{1}{{\sqrt V }}\sum\nolimits_\vec{k} e^{ - i\vec{k}\cdot \vec{j}}c_{\vec{k}\sigma } $ leads to
\begin{eqnarray}
S_{\delta n}(t) = \frac{1}{ V^2}\sum\limits_{(\vec{j},\vec{j}' \in A),\sigma ,\sigma '} {\sum\limits_{{\vec{k}_1},{\vec{k}_2},{\vec{k}_3},{\vec{k}_4}} {{e^{i\left[ {\left( {{\vec{k}_1} - {\vec{k}_2}} \right).\vec{j} + \left( {{\vec{k}_3} - {\vec{k}_4}} \right).\vec{j}'} \right]}}} \left\langle {c_{{\vec{k}_1}\sigma }^\dag (t){c_{{\vec{k}_4}\sigma '}}(0)} \right\rangle \left\langle {{c_{{\vec{k}_2}\sigma }}(t)c_{{\vec{k}_3}\sigma '}^\dag (0)} \right\rangle } 
\end{eqnarray}
Using the transformation $c_{\vec{k}\sigma } = \sum\nolimits_m \langle {\sigma |{\omega _{\vec{k}m}}} \rangle d_{\vec{k}m}$,  we can go on the diagonal basis of the Hamiltonian where the time evolved annihilation operator reads 
\begin{eqnarray}
{c_{\vec{k}\sigma }}(t) = \sum\nolimits_m {\left\langle {\sigma |{\omega _{\vec{k}m}}} \right\rangle {d_{\vec{k}m}}} (t) = \sum\nolimits_m {{e^{ - i{\omega _{\vec{k}m}}t}}\left\langle {\sigma |{\omega _{\vec{k}m}}} \right\rangle {d_{\vec{k}m}}}. 
\end{eqnarray}
Using the Fourier form and creation/annihilation operators in the Hamiltonian diagonal basis, the temporal correlation of fluctuations becomes
\begin{eqnarray}
{S_{\delta n}}(t) 
 &=& \frac{1}{V^2}\sum\limits_{(\vec{j},\vec{j}' \in A),\sigma ,\sigma '} \sum\limits_{\vec{k}_1,\vec{k}_2,\vec{k}_3,\vec{k}_4} \sum\limits_{m_1,m_2,m_3,m_4} e^{i\left(\vec{k}_1 - \vec{k}_2\right).\vec{j} + i\left(\vec{k}_3 - \vec{k}_4 \right).\vec{j}' }  e^{i(\omega _{\vec{k}_1m_1} - \omega _{\vec{k}_2m_2})t} \nonumber\\
 &\times& \langle {{\omega _{{\vec{k}_1}{m_1}}}|\sigma } \rangle \langle {\sigma '|{\omega _{{\vec{k}_4}{m_4}}}} \rangle \langle {\sigma |{\omega _{{\vec{k}_2}{m_2}}}} \rangle \langle {{\omega _{{\vec{k}_3}{m_3}}}|\sigma '} \rangle \langle d_{\vec{k}_1m_1}^\dag d_{\vec{k}_4m_4} \rangle \langle d_{\vec{k}_2m_2}d_{\vec{k}_3m_3}^\dag \rangle \nonumber\\
  &=& \frac{1}{V^2}\sum\limits_{(\vec{j},\vec{j}' \in A)} \sum\limits_{\vec{k}_1,\vec{k}_2} \sum\limits_{m_1,m_2} e^{i\left(\vec{k}_1 - \vec{k}_2\right).(\vec{j}-\vec{j}')} e^{i(\omega _{\vec{k}_1m_1} - \omega _{\vec{k}_2m_2})t} \nonumber\\
 &\times& 
 {\rm Tr}\left[
 \ket{\omega_{{\vec{k}_1}{m_1}}}
 \langle {\omega_{{\vec{k}_1}{m_1}}}|
  \ket{\omega _{{\vec{k}_2}{m_2}}}
 \langle {\omega _{{\vec{k}_2}{m_2}}}| \right]
n(\omega _{\vec{k}_1m_1}) [1-n(\omega _{\vec{k}_2m_2})].
\end{eqnarray}
In the second expression, we have substituted the expectation values in terms of the Fermi-Dirac (FD) distribution $n(\omega _{\vec{k}m})=(1+e^{\beta \omega_{\vec{k}m}})^{-1}$ as
\begin{eqnarray}
\langle d_{{\vec{k}_1}{m_1}}^\dag d_{{\vec{k}_4}{m_4}}\rangle  &=& n({\omega _{{\vec{k}_1}{m_1}}}){\delta _{{\vec{k}_1}{\vec{k}_4}}}{\delta _{{m_1}{m_4}}}  \nonumber\\
\langle d_{\vec{k}_2m_2}d_{\vec{k}_3m_3}^\dag  \rangle  &=&  \delta _{\vec{k}_2\vec{k}_3}\delta _{m_2m_3} - \langle d_{\vec{k}_3m_3}^\dag d_{\vec{k}_2m_2} \rangle   = \left[ 1 - n(\omega _{\vec{k}_3m_3}) \right]\delta _{\vec{k}_2\vec{k}_3}\delta _{m_2m_3},
\end{eqnarray}
and carried out the summation over internal degrees of freedom $\sigma$ and $\sigma'$. Moreover, we define band projectors $P_{\vec{k}m}=|\omega_{\vec{k}m}\rangle \langle \omega_{\vec{k}m}|$ which will be used in the following for the sake of compactness.
At zero temperature, which is the main focus in this work, the FD distribution reduces to a step function form $n(\omega)=\Theta(\omega)$ where the chemical potential $\mu$ is absorbed in the dispersion relations.

Now, by taking the Fourier transform, we get the dynamical structure factor as
\begin{eqnarray}
S_{\delta n}(\omega) &=& \int_{-\infty}^{\infty}dt e^{i\omega t} S_{\delta n}(t) \nonumber\\
&=&\frac{1}{V^2}\sum\limits_{(\vec{j},\vec{j}' \in A) } \sum\limits_{\vec{k}_1,\vec{k}_2} \sum\limits_{m_1,m_2} e^{i(\vec{k}_1 - \vec{k}_2).(\vec{j}-\vec{j}')} 
{\rm Tr}\left( P_{\vec{k}_1m_1}\: P_{\vec{k}_2m_2} \right)
 n(\omega _{\vec{k}_1m_1}) [1-n(\omega _{\vec{k}_2m_2})] 
 \int_{-\infty}^{\infty}dt e^{i( \omega+\omega _{\vec{k}_1m_1} - \omega _{\vec{k}_2m_2} )t} 
 \nonumber\\
&=&\frac{1}{V^2}\sum\limits_{(\vec{j},\vec{j}' \in A) } \sum\limits_{\vec{k}_1,\vec{k}_2} \sum\limits_{m_1,m_2} e^{i(\vec{k}_1 - \vec{k}_2).(\vec{j}-\vec{j}')} 
{\rm Tr}\left( P_{\vec{k}_1m_1}\: P_{\vec{k}_2m_2} \right)
 n(\omega _{\vec{k}_1m_1}) [1-n(\omega _{\vec{k}_2m_2})] 
 \delta ( \omega+\omega _{\vec{k}_1m_1} - \omega _{\vec{k}_2m_2} )
\end{eqnarray}

\end{widetext}

\bibliography{refs.bib}

\begin{thebibliography}{40}
\bibitem [{foo()}]{footnote:computation}%
  \BibitemOpen
  \bibinfo {note} {Although the expression in Eq.~\eqref{eq:noise2} can be
  evaluated straightforwardly, the fourfold summation scales rapidly with
  system size. Consequently, the results presented here are limited to lattices
  with $N_x = N_y \sim 20$.}\BibitemShut {Stop}%
\end{thebibliography}

\end{document}